\documentclass[12pt,reqno]{amsart}
\usepackage{graphicx}
\usepackage{amscd}
\usepackage{amsmath}
\usepackage{epsfig}
\usepackage{amsfonts}
\usepackage{amssymb}

\providecommand{\U}[1]{\protect\rule{.1in}{.1in}}
\providecommand{\U}[1]{\protect\rule{.1in}{.1in}}
\allowdisplaybreaks

\theoremstyle{plain}

\numberwithin{equation}{section}

\input{tcilatex}

\begin{document}
\title[ Relativistic Coulomb Integrals]{Mathematical Structure of
Relativistic Coulomb Integrals}
\author{Sergei K. Suslov}
\address{School of Mathematical and Statistical Sciences and Mathematical,
Computational and Modeling Sciences Center, Arizona State University, Tempe,
AZ 85287--1804, U.S.A.}
\email{sks@asu.edu}
\urladdr{http://hahn.la.asu.edu/\symbol{126}suslov/index.html}
\date{\today }

\begin{abstract}
We show that the diagonal matrix elements $\langle Or^{p}\rangle ,$ where $O$
$=\left\{ 1,\beta ,i\mathbf{\alpha n}\beta \right\} $ are the standard Dirac
matrix operators and the angular brackets denote the quantum-mechanical
average for the relativistic Coulomb problem, may be considered as
difference analogs of the radial wave functions. Such structure provides an
independent way of obtaining closed forms of these matrix elements by
elementary methods of the theory of difference equations without explicit
evaluation of the integrals. Three-term recurrence relations for each of
these expectation values are derived as a by-product. Transformation
formulas for the corresponding generalized hypergeometric series are
discussed.
\end{abstract}

\maketitle

\section{Introduction}

Recent experimental and theoretical progress has renewed interest in quantum
electrodynamics of atomic hydrogenlike systems (see, for example, \cite%
{Gum05}, \cite{Gum07}, \cite{Karsh01}, \cite{Kash03}, \cite{Mohr:Plun:Soff98}%
, \cite{ShabGreen}, and \cite{ShabYFN08} and references therein). In the
last decade, the two-time Green's function method of deriving formal
expressions for the energy shift of a bound-state level of high-$Z$
few-electron systems was developed \cite{ShabGreen} and numerical
calculations of QED effects in heavy ions were performed with excellent
agreement to current experimental data \cite{Gum05}, \cite{Gum07}, \cite%
{ShabYFN08}. These advances motivate detailed study of the expectation
values of the Dirac matrix operators between the bound-state relativistic
Coulomb wave functions. Special cases appear in calculations of the magnetic
dipole hyperfine splitting, the electric quadrupole hyperfine splitting, the
anomalous Zeeman effect, and the relativistic recoil corrections in
hydrogenlike ions (see, for example, \cite{Adkins}, \cite{ShabHyd}, \cite%
{ShabHydVir}, \cite{Suslov} and references therein). These expectation
values can be used in calculations with hydrogenlike wave functions when a
high precision is required.\medskip

In the previous paper \cite{Suslov}, we have evaluated the relativistic
Coulomb integrals of the radial functions,%
\begin{eqnarray}
A_{p} &=&\int_{0}^{\infty }r^{p+2}\left( F^{2}\left( r\right) +G^{2}\left(
r\right) \right) \ dr,  \label{meA} \\
B_{p} &=&\int_{0}^{\infty }r^{p+2}\left( F^{2}\left( r\right) -G^{2}\left(
r\right) \right) \ dr,  \label{meB} \\
C_{p} &=&\int_{0}^{\infty }r^{p+2}F\left( r\right) G\left( r\right) \ dr,
\label{meC}
\end{eqnarray}%
for all admissible powers $p,$ in terms of three special generalized
hypergeometric $_{3}F_{2}$ series related to the Chebyshev polynomials of a
discrete variable \cite{Ni:Su:Uv} (we concentrate on the radial integrals
since, for problems involving spherical symmetry, one can reduce all
expectation values to radial integrals by use of the properties of angular
momentum). These integrals are linearly dependent: 
\begin{equation}
\left( 2\kappa +\varepsilon \left( p+1\right) \right) A_{p}-\left(
2\varepsilon \kappa +p+1\right) B_{p}=4\mu C_{p}  \label{indint1}
\end{equation}%
(see, for example, \cite{Adkins}, \cite{ShabVest}, \cite{Shab91}, and \cite%
{Suslov} for more details). Thus, eliminating, say $C_{p},$ one can deal
with $A_{p}$ and $B_{p}$ only. The corresponding representations in terms of
only two linearly independent generalized hypergeometric series are given in
this paper (see (\ref{TA})--(\ref{TC}) and (\ref{NUA})--(\ref{NUC})).\medskip

The integrals (\ref{meA})--(\ref{meC}) satisfy numerous recurrence relations
in $p,$ which provide an effective way of their evaluation for small $p$
(see \cite{Adkins}, \cite{ShabVest}, \cite{Shab91}, \cite{Suslov} and
references therein). The two-term recurrence relations were derived by
Shabaev \cite{ShabVest}, \cite{Shab91} on the basis of a hypervirial theorem
and by a different method using relativistic versions of the
Kramers--Pasternack three-term recurrence relations in \cite{SuslovB}. In
our notations,%
\begin{eqnarray}
A_{p+1} &=&-\left( p+1\right) \frac{4\nu ^{2}\varepsilon +2\kappa \left(
p+2\right) +\varepsilon \left( p+1\right) \left( 2\kappa \varepsilon
+p+2\right) }{4\left( 1-\varepsilon ^{2}\right) \left( p+2\right) \beta \mu }%
\ A_{p}  \label{rra} \\
&&+\frac{4\mu ^{2}\left( p+2\right) +\left( p+1\right) \left( 2\kappa
\varepsilon +p+1\right) \left( 2\kappa \varepsilon +p+2\right) }{4\left(
1-\varepsilon ^{2}\right) \left( p+2\right) \beta \mu }\ B_{p},  \notag
\end{eqnarray}%
\begin{eqnarray}
B_{p+1} &=&-\left( p+1\right) \frac{4\nu ^{2}+2\kappa \varepsilon \left(
2p+3\right) +\varepsilon ^{2}\left( p+1\right) \left( p+2\right) }{4\left(
1-\varepsilon ^{2}\right) \left( p+2\right) \beta \mu }\ A_{p}  \label{rrb}
\\
&&+\frac{4\mu ^{2}\varepsilon \left( p+2\right) +\left( p+1\right) \left(
2\kappa \varepsilon +p+1\right) \left( 2\kappa +\varepsilon \left(
p+2\right) \right) }{4\left( 1-\varepsilon ^{2}\right) \left( p+2\right)
\beta \mu }\ B_{p}  \notag
\end{eqnarray}%
and%
\begin{eqnarray}
A_{p-1} &=&\beta \frac{4\mu ^{2}\varepsilon \left( p+1\right) +p\left(
2\kappa \varepsilon +p\right) \left( 2\kappa +\varepsilon \left( p+1\right)
\right) }{\mu \left( 4\nu ^{2}-p^{2}\right) p}\ A_{p}  \label{rrab} \\
&&-\beta \frac{4\mu ^{2}\left( p+1\right) +p\left( 2\kappa \varepsilon
+p\right) \left( 2\kappa \varepsilon +p+1\right) }{\mu \left( 4\nu
^{2}-p^{2}\right) p}\ B_{p},  \notag
\end{eqnarray}%
\begin{eqnarray}
B_{p-1} &=&\beta \frac{4\nu ^{2}+2\kappa \varepsilon \left( 2p+1\right)
+\varepsilon ^{2}p\left( p+1\right) }{\mu \left( 4\nu ^{2}-p^{2}\right) }\
A_{p}  \label{rrba} \\
&&-\beta \frac{4\nu ^{2}\varepsilon +2\kappa \left( p+1\right) +\varepsilon
p\left( 2\kappa \varepsilon +p+1\right) }{\mu \left( 4\nu ^{2}-p^{2}\right) }%
\ B_{p},  \notag
\end{eqnarray}%
respectively. Here,%
\begin{eqnarray}
&&\kappa =\pm \left( j+1/2\right) ,\qquad \nu =\sqrt{\kappa ^{2}-\mu ^{2}}, 
\notag \\
&&\mu =\alpha Z=Ze^{2}/\hbar c,\qquad a=\sqrt{1-\varepsilon ^{2}},
\label{notations} \\
&&\varepsilon =E/mc^{2},\qquad \beta =mc/\hbar  \notag
\end{eqnarray}%
with the total angular momentum $j=1/2,3/2,5/2,\ ...$ (see \cite{Suslov} and 
\cite{Sus:Trey} for more details).\medskip

These recurrence relations are complemented by the symmetries of the
integrals $A_{p},$ $B_{p},$ and $C_{p}$ under reflections $p\rightarrow -p-1$
and $p\rightarrow -p-3$ found in \cite{Suslov} (see also \cite{Andrae97}).
For example,%
\begin{eqnarray}
A_{-p-3} &=&\left( 2a\beta \right) ^{2p+3}\ \frac{\Gamma \left( 2\nu
-p-2\right) }{\Gamma \left( 2\nu +p+3\right) }  \label{invA} \\
&&\times \left( -\left( p+1\right) \frac{4\nu ^{2}+2\varepsilon \kappa
\left( 2p+3\right) -\left( p+2\right) ^{2}}{p+2}\ A_{p}\right.  \notag \\
&&\qquad +\left. 2\kappa \left( 2\varepsilon \kappa -1\right) \frac{2p+3^{\ }%
}{p+2}\ B_{p}\right) ,  \notag
\end{eqnarray}%
\begin{eqnarray}
B_{-p-3} &=&\left( 2a\beta \right) ^{2p+3}\ \frac{\Gamma \left( 2\nu
-p-2\right) }{\Gamma \left( 2\nu +p+3\right) }  \label{invB} \\
&&\times \left( \varepsilon \left( p+1\right) \left( 2p+3\right) \
A_{p}+\left( 4\nu ^{2}-2\varepsilon \kappa \left( 2p+3\right) -\left(
p+1\right) ^{2}\right) \ B_{p}\right) ,  \notag
\end{eqnarray}%
for independent convergent integrals $A_{p}$ and $B_{p}.$\medskip

In this paper, we would like to draw reader's attention to an interesting
analogy between the explicit solutions of the first order system of
difference equations (\ref{rra})--(\ref{rrb}) and the standard method of
dealing with the system of differential equations for the radial
relativistic Coulomb wave functions $F$ and $G$ (see, for example, \cite%
{Be:Sal}, \cite{Dar}, \cite{Gor}, \cite{Ni:Uv}, \cite{Schiff}, and \cite%
{Sus:Trey} regarding solution of the Dirac equation in Coulomb field). En
route, we derive the three-term recurrence relations for each of the single
integrals (\ref{meA})--(\ref{meC}) that seem to be new and convenient for
their evaluation. Our observation provides an independent method of
obtaining closed forms of these matrix elements, but this time, from the
theory of difference equations and without explicit evaluation of the
integrals. Some transformation formulas for the corresponding generalized
hypergeometric series are derived.

\section{Three-term Recurrence Relations}

Several relativistic Kramers--Pasternack three-term vector recurrence
relations for the integrals $A_{p},$ $B_{p},$ $C_{p}$ have been obtained in 
\cite{SuslovB}. A more general setting is as follows. Let us rewrite (\ref%
{rra})--(\ref{rrb}) and (\ref{rrab})--(\ref{rrba}) in the matrix form%
\begin{equation}
\left( 
\begin{array}{c}
A_{p}\medskip \\ 
B_{p}%
\end{array}%
\right) =S_{p}\left( 
\begin{array}{c}
A_{p-1}\medskip \\ 
B_{p-1}%
\end{array}%
\right) ,\qquad \left( 
\begin{array}{c}
A_{p-1}\medskip \\ 
B_{p-1}%
\end{array}%
\right) =S_{p}^{-1}\left( 
\begin{array}{c}
A_{p}\medskip \\ 
B_{p}%
\end{array}%
\right)  \label{MatSol}
\end{equation}%
and denote%
\begin{equation}
S_{p}=\left( 
\begin{array}{cc}
a_{p} & b_{p}\medskip \\ 
c_{p} & d_{p}%
\end{array}%
\right) ,\qquad S_{p}^{-1}=\frac{1}{\Delta _{p}}\left( 
\begin{array}{cc}
d_{p} & -b_{p}\medskip \\ 
-c_{p} & a_{p}%
\end{array}%
\right)  \label{Smat}
\end{equation}%
with%
\begin{eqnarray}
a_{p} &=&-p\frac{4\nu ^{2}\varepsilon +2\kappa \left( p+1\right)
+\varepsilon p\left( 2\kappa \varepsilon +p+1\right) }{4\left( 1-\varepsilon
^{2}\right) \left( p+1\right) \beta \mu },  \label{ap} \\
b_{p}\medskip &=&\frac{4\mu ^{2}\left( p+1\right) +p\left( 2\kappa
\varepsilon +p\right) \left( 2\kappa \varepsilon +p+1\right) }{4\left(
1-\varepsilon ^{2}\right) \left( p+1\right) \beta \mu },  \label{bp} \\
c_{p} &=&-p\frac{4\nu ^{2}+2\kappa \varepsilon \left( 2p+1\right)
+\varepsilon ^{2}p\left( p+1\right) }{4\left( 1-\varepsilon ^{2}\right)
\left( p+1\right) \beta \mu },  \label{cp} \\
d_{p} &=&\frac{4\mu ^{2}\varepsilon \left( p+1\right) +p\left( 2\kappa
\varepsilon +p\right) \left( 2\kappa +\varepsilon \left( p+1\right) \right) 
}{4\left( 1-\varepsilon ^{2}\right) \left( p+1\right) \beta \mu }  \label{dp}
\end{eqnarray}%
and%
\begin{equation}
\Delta _{p}=\det S_{p}=\frac{\left( 4\nu ^{2}-p^{2}\right) p}{\left( 2a\beta
\right) ^{2}\left( p+1\right) }.  \label{detS}
\end{equation}%
Eliminating $A_{p}$ and $B_{p},$ respectively, from the system (\ref{MatSol}%
), we arrive at the following three-term recurrence equations for the
independent integrals%
\begin{eqnarray}
A_{p+1} &=&\left( a_{p+1}+\frac{b_{p+1}}{b_{p}}d_{p}\right) \ A_{p}-\frac{%
b_{p+1}}{b_{p}}\Delta _{p}\ A_{p-1},  \label{3termA} \\
B_{p+1} &=&\left( d_{p+1}+\frac{c_{p+1}}{c_{p}}a_{p}\right) \ B_{p}-\frac{%
c_{p+1}}{c_{p}}\Delta _{p}\ B_{p-1},  \label{3termB}
\end{eqnarray}%
which seem are missing in the available literature.\medskip

In general, one can easily verify that the following vector three-term
recurrence relation holds:%
\begin{equation}
\left( 
\begin{array}{c}
A_{p+1}\medskip  \\ 
B_{p+1}%
\end{array}%
\right) =M_{p}\left( 
\begin{array}{c}
A_{p}\medskip  \\ 
B_{p}%
\end{array}%
\right) +N_{p}\left( 
\begin{array}{c}
A_{p-1}\medskip  \\ 
B_{p-1}%
\end{array}%
\right)   \label{3termGen}
\end{equation}%
for two matrices $M_{p}$ and $N_{p}$ provided that%
\begin{equation}
S_{p+1}=M_{p}+N_{p}S_{p}^{-1}.  \label{3termMat}
\end{equation}%
Our equations (\ref{3termA})--(\ref{3termB}) provide a diagonal matrix
solution. According to (\ref{MatSol}), (\ref{3termGen}) and (\ref{3termMat}%
), a simple identity%
\begin{equation}
\left( 
\begin{array}{c}
A_{p+1}\medskip  \\ 
B_{p+1}%
\end{array}%
\right) =\left( S_{p+1}-N_{p}S_{p}^{-1}\right) \left( 
\begin{array}{c}
A_{p}\medskip  \\ 
B_{p}%
\end{array}%
\right) +N_{p}\left( 
\begin{array}{c}
A_{p-1}\medskip  \\ 
B_{p-1}%
\end{array}%
\right)   \label{3termSol}
\end{equation}%
holds for any matrix $N_{p}.$ The known three-term recurrence relations for
the relativistic Coulomb integrals can be obtained by choosing different
forms of the matrix $N_{p}.$ The case $N_{p}=0$ goes back to the two-term
recurrence relation (\ref{MatSol}) and two more explicit solutions have been
found in \cite{SuslovB}. Here we analyze another possibility and take%
\begin{equation*}
N_{p}=\left( 
\begin{array}{cc}
\lambda _{p} & 0\medskip  \\ 
0 & \mu _{p}%
\end{array}%
\right) \quad \text{and\quad }N_{p}=\left( 
\begin{array}{cc}
0 & \lambda _{p}\medskip  \\ 
\mu _{p} & 0%
\end{array}%
\right) 
\end{equation*}%
for suitable parameters $\lambda _{p}$ and $\mu _{p}.$ A new convenient
relations are as follows%
\begin{eqnarray}
\left( 
\begin{array}{c}
A_{p+1}\medskip  \\ 
B_{p+1}%
\end{array}%
\right)  &=&\left( 
\begin{array}{cc}
a_{p+1}+b_{p+1}\dfrac{d_{p}}{b_{p}} & 0\medskip  \\ 
0 & d_{p+1}+c_{p+1}\dfrac{a_{p}}{c_{p}}%
\end{array}%
\right) \left( 
\begin{array}{c}
A_{p}\medskip  \\ 
B_{p}%
\end{array}%
\right)   \notag \\
&&-\Delta _{p}\left( 
\begin{array}{cc}
b_{p+1}/b_{p} & 0\medskip  \\ 
0 & c_{p+1}/c_{p}%
\end{array}%
\right) \left( 
\begin{array}{c}
A_{p-1}\medskip  \\ 
B_{p-1}%
\end{array}%
\right) ,  \label{Mat1}
\end{eqnarray}%
\begin{eqnarray}
\left( 
\begin{array}{c}
A_{p+1}\medskip  \\ 
B_{p+1}%
\end{array}%
\right)  &=&\left( 
\begin{array}{cc}
a_{p+1}+b_{p+1}\dfrac{c_{p}}{a_{p}} & 0\medskip  \\ 
0 & d_{p+1}+c_{p+1}\dfrac{b_{p}}{d_{p}}%
\end{array}%
\right) \left( 
\begin{array}{c}
A_{p}\medskip  \\ 
B_{p}%
\end{array}%
\right)   \notag \\
&&+\Delta _{p}\left( 
\begin{array}{cc}
0 & b_{p+1}/a_{p}\medskip  \\ 
c_{p+1}/d_{p} & 0%
\end{array}%
\right) \left( 
\begin{array}{c}
A_{p-1}\medskip  \\ 
B_{p-1}%
\end{array}%
\right) ,  \label{Mat2}
\end{eqnarray}%
\begin{eqnarray}
\left( 
\begin{array}{c}
A_{p+1}\medskip  \\ 
B_{p+1}%
\end{array}%
\right)  &=&\left( 
\begin{array}{cc}
0 & b_{p+1}+a_{p+1}\dfrac{b_{p}}{d_{p}}\medskip  \\ 
c_{p+1}+d_{p+1}\dfrac{c_{p}}{a_{p}} & 0%
\end{array}%
\right) \left( 
\begin{array}{c}
A_{p}\medskip  \\ 
B_{p}%
\end{array}%
\right)   \notag \\
&&+\Delta _{p}\left( 
\begin{array}{cc}
a_{p+1}/d_{p} & 0\medskip  \\ 
0 & d_{p+1}/a_{p}%
\end{array}%
\right) \left( 
\begin{array}{c}
A_{p-1}\medskip  \\ 
B_{p-1}%
\end{array}%
\right) ,  \label{Mat3}
\end{eqnarray}%
\begin{eqnarray}
\left( 
\begin{array}{c}
A_{p+1}\medskip  \\ 
B_{p+1}%
\end{array}%
\right)  &=&\left( 
\begin{array}{cc}
0 & b_{p+1}+a_{p+1}\dfrac{a_{p}}{c_{p}}\medskip  \\ 
c_{p+1}+d_{p+1}\dfrac{d_{p}}{b_{p}} & 0%
\end{array}%
\right) \left( 
\begin{array}{c}
A_{p}\medskip  \\ 
B_{p}%
\end{array}%
\right)   \notag \\
&&-\Delta _{p}\left( 
\begin{array}{cc}
0 & a_{p+1}/c_{p}\medskip  \\ 
d_{p+1}/b_{p} & 0%
\end{array}%
\right) \left( 
\begin{array}{c}
A_{p-1}\medskip  \\ 
B_{p-1}%
\end{array}%
\right) .  \label{Mat4}
\end{eqnarray}%
Explicit diagonal form, when the equations are separated, is given by%
\begin{eqnarray}
A_{p+1} &=&\frac{\mu P\left( p\right) }{a^{2}\beta \left( 4\mu ^{2}\left(
p+1\right) +p\left( 2\kappa \varepsilon +p\right) \left( 2\kappa \varepsilon
+p+1\right) \right) \left( p+2\right) }\ A_{p}  \label{3termAA} \\
&&-\frac{\left( 4\nu ^{2}-p^{2}\right) \left( 4\mu ^{2}\left( p+2\right)
+\left( p+1\right) \left( 2\kappa \varepsilon +p+1\right) \left( 2\kappa
\varepsilon +p+2\right) \right) p}{\left( 2a\beta \right) ^{2}\left( 4\mu
^{2}\left( p+1\right) +p\left( 2\kappa \varepsilon +p\right) \left( 2\kappa
\varepsilon +p+1\right) \right) \left( p+2\right) }\ A_{p-1},  \notag
\end{eqnarray}%
\begin{eqnarray}
B_{p+1} &=&\frac{\varepsilon \mu Q\left( p\right) }{a^{2}\beta \left( 4\nu
^{2}+2\kappa \varepsilon \left( 2p+1\right) +\varepsilon ^{2}p\left(
p+1\right) \right) \left( p+2\right) }\ B_{p}  \label{3termBB} \\
&&-\frac{\left( 4\nu ^{2}-p^{2}\right) \left( 4\nu ^{2}+2\kappa \varepsilon
\left( 2p+3\right) +\varepsilon ^{2}\left( p+1\right) \left( p+2\right)
\right) \left( p+1\right) }{\left( 2a\beta \right) ^{2}\left( 4\nu
^{2}+2\kappa \varepsilon \left( 2p+1\right) +\varepsilon ^{2}p\left(
p+1\right) \right) \left( p+2\right) }\ B_{p-1},  \notag
\end{eqnarray}%
where%
\begin{eqnarray}
P\left( p\right)  &=&\allowbreak 2\varepsilon p^{4}+\left( 8\kappa
\varepsilon ^{2}+5\varepsilon \right) p^{3}  \label{PolP} \\
&&+\left( 8\kappa ^{2}\varepsilon ^{3}+8\kappa ^{2}\varepsilon +20\kappa
\varepsilon ^{2}-4\kappa -8\nu ^{2}\varepsilon +3\varepsilon \right)
\allowbreak p^{2}  \notag \\
&&+\left( 12\kappa ^{2}\varepsilon ^{3}+20\kappa ^{2}\varepsilon +16\kappa
\varepsilon ^{2}-10\kappa -20\nu ^{2}\varepsilon \right) \allowbreak p 
\notag \\
&&+4\kappa ^{2}\varepsilon ^{3}+8\kappa ^{2}\varepsilon +4\kappa \varepsilon
^{2}-4\kappa -12\nu ^{2}\varepsilon ,  \notag \\
Q\left( p\right)  &=&2\varepsilon ^{2}p^{3}+\left( 7\varepsilon ^{2}+8\kappa
\varepsilon -2\right) p^{2}  \label{PolQ} \\
&&+\left( 8\nu ^{2}+7\varepsilon ^{2}+16\kappa \varepsilon -4\right)
\allowbreak p  \notag \\
&&+12\nu ^{2}+2\varepsilon ^{2}+6\kappa \varepsilon -2.  \notag
\end{eqnarray}%
In comparison with other papers (see \cite{Adkins}, \cite{Andrae97}, \cite%
{ShabVest}, \cite{Shab91}, \cite{Suslov}, \cite{SuslovB}, and references
therein), our consideration provides an alternative way of the recursive
evaluation of the special values $A_{p}$ and $B_{p},$ when one deals
separately with one of these integrals only. The corresponding initial data
can be found in \cite{Suslov}. It is important emphasizing, for the purpose
of this paper, that this argument resembles the reduction of the first order
system of differential equations for relativistic radial Coulomb wave
functions $F$ and $G$ to the second order differential equations (see, for
example, \cite{Ni:Uv} and \cite{Sus:Trey}).\medskip 

If one wants to solve equations (\ref{3termAA})--(\ref{3termBB})
analytically for all admissible powers, then the major obstacle is that they
are not difference equations of hypergeometric type on a quadratic lattice,
solutions of which are available in the literature \cite{Ni:Su:Uv}. The
following consideration helps. A linear transformation%
\begin{equation}
\left( 
\begin{array}{c}
X_{p}\medskip \\ 
Y_{p}%
\end{array}%
\right) =T_{p}\ \left( 
\begin{array}{c}
A_{p}\medskip \\ 
B_{p}%
\end{array}%
\right) ,  \label{Trans}
\end{equation}%
where%
\begin{equation}
T_{p}=\left( 
\begin{array}{cc}
\alpha _{p} & \beta _{p}\medskip \\ 
\gamma _{p} & \delta _{p}%
\end{array}%
\right) ,\qquad \det T_{p}=\alpha _{p}\delta _{p}-\beta _{p}\gamma _{p}\neq
0,  \label{TransT}
\end{equation}%
results in a new system of the first order difference equations%
\begin{equation}
\left( 
\begin{array}{c}
X_{p}\medskip \\ 
Y_{p}%
\end{array}%
\right) =\widetilde{S}_{p}\left( 
\begin{array}{c}
X_{p-1}\medskip \\ 
Y_{p-1}%
\end{array}%
\right) ,  \label{MatSolX}
\end{equation}%
where the corresponding similar matrix is given by%
\begin{equation}
\widetilde{S}_{p}=T_{p}S_{p}T_{p-1}^{-1}=\left( 
\begin{array}{cc}
\widetilde{a}_{p} & \widetilde{b}\medskip _{p} \\ 
\widetilde{c}_{p} & \widetilde{d}_{p}%
\end{array}%
\right)  \label{NewS}
\end{equation}%
with%
\begin{eqnarray}
\det T_{p-1}\ \widetilde{a}_{p} &=&\alpha _{p}\delta _{p-1}a_{p}-\alpha
_{p}\gamma _{p-1}b_{p}+\beta _{p}\delta _{p-1}c_{p}-\beta _{p}\gamma
_{p-1}d_{p},  \label{T1} \\
\det T_{p-1}\ \widetilde{b}\medskip _{p} &=&-\alpha _{p}\beta
_{p-1}a_{p}+\alpha _{p}\alpha _{p-1}b_{p}-\beta _{p}\beta _{p-1}c_{p}+\beta
_{p}\alpha _{p-1}d_{p},  \label{T2} \\
\det T_{p-1}\ \widetilde{c}_{p} &=&\gamma _{p}\delta _{p-1}a_{p}-\gamma
_{p}\gamma _{p-1}b_{p}+\delta _{p}\delta _{p-1}c_{p}-\delta _{p}\gamma
_{p-1}d_{p},  \label{T3} \\
\det T_{p-1}\ \widetilde{d}_{p} &=&-\gamma _{p}\beta _{p-1}a_{p}+\gamma
_{p}\alpha _{p-1}b_{p}-\delta _{p}\beta _{p-1}c_{p}+\delta _{p}\alpha
_{p-1}d_{p},  \label{T4}
\end{eqnarray}%
and%
\begin{equation}
\widetilde{\Delta }_{p}=\det \widetilde{S}_{p}=\det S_{p}\ \frac{\det T_{p}}{%
\det T_{p-1}}\neq 0.
\end{equation}%
The new separated three-term recurrence equations take the similar forms%
\begin{eqnarray}
X_{p+1} &=&\left( \widetilde{a}_{p+1}+\frac{\widetilde{b}_{p+1}}{\widetilde{b%
}_{p}}\widetilde{d}_{p}\right) \ X_{p}-\frac{\widetilde{b}_{p+1}}{\widetilde{%
b}_{p}}\widetilde{\Delta }_{p}\ X_{p-1},  \label{3termX} \\
Y_{p+1} &=&\left( \widetilde{d}_{p+1}+\frac{\widetilde{c}_{p+1}}{\widetilde{c%
}_{p}}\widetilde{a}_{p}\right) \ Y_{p}-\frac{\widetilde{c}_{p+1}}{\widetilde{%
c}_{p}}\widetilde{\Delta }_{p}\ Y_{p-1}.  \label{3termY}
\end{eqnarray}%
As in the case of the radial wave functions \cite{Ni:Uv} and \cite{Sus:Trey}%
, there are several possibilities to choose the matrix $T_{p}$ in order to
simplify the original equations (\ref{3termAA})--(\ref{3termBB}). Examples
of such transformations, when the resulting equations are of a
hypergeometric type and coincide with difference equations for special dual
Hahn polynomials \cite{Karlin:McGregor61}, \cite{Ko:Sw}, \cite{Ni:Su:Uv}
(see also appendix~A), are given in the next section.

\section{Transformations of Relativistic Coulomb Integrals}

The integrals $A_{p},$ $B_{p},$ and $C_{p}$ can be evaluated in terms of two
linearly independent $_{3}F_{2}$ functions, which are related to the special
dual Hahn polynomials that can be thought of as difference analogs of the
Laguerre polynomials in explicit formulas for the radial wave functions (see 
\cite{Ni:Uv} and \cite{Sus:Trey} for a detailed tutorial on solution of the
relativistic Coulomb problem). This fact has been partially explored in \cite%
{Suslov} and we elaborate on this connection here. Two different
representations of the expectation values are available in a complete
analogy with the well-known structure of the relativistic wave
functions.\medskip

Analogs of the traditional forms are as follows%
\begin{eqnarray}
&&2\left( p+1\right) a\mu \left( 2a\beta \right) ^{p}\ \frac{\Gamma \left(
2\nu +1\right) }{\Gamma \left( 2\nu +p+1\right) }\ A_{p}  \label{TA} \\
&&\qquad =\left( \mu +a\kappa \right) \left( a\left( 2\varepsilon \kappa
+p+1\right) -2\varepsilon \mu \right)  \notag \\
&&\qquad \quad \times ~_{3}F_{2}\left( 
\begin{array}{c}
1-n,\ -p,\ p+1\medskip \\ 
2\nu +1,\quad 1%
\end{array}%
\right)  \notag \\
&&\qquad +~\left( \mu -a\kappa \right) \left( a\left( 2\varepsilon \kappa
+p+1\right) +2\varepsilon \mu \right)  \notag \\
&&\qquad \qquad \times ~_{3}F_{2}\left( 
\begin{array}{c}
-n,\ -p,\ p+1\medskip \\ 
2\nu +1,\quad 1%
\end{array}%
\right) ,  \notag
\end{eqnarray}%
\begin{eqnarray}
&&2\left( p+1\right) a\mu \left( 2a\beta \right) ^{p}\ \frac{\Gamma \left(
2\nu +1\right) }{\Gamma \left( 2\nu +p+1\right) }\ B_{p}  \label{TB} \\
&&\qquad =\left( \mu +a\kappa \right) \left( a\left( 2\kappa +\varepsilon
\left( p+1\right) \right) -2\mu \right)  \notag \\
&&\qquad \quad \times ~_{3}F_{2}\left( 
\begin{array}{c}
1-n,\ -p,\ p+1\medskip \\ 
2\nu +1,\quad 1%
\end{array}%
\right)  \notag \\
&&\qquad +~\left( \mu -a\kappa \right) \left( a\left( 2\kappa +\varepsilon
\left( p+1\right) \right) +2\mu \right)  \notag \\
&&\qquad \qquad \times ~_{3}F_{2}\left( 
\begin{array}{c}
-n,\ -p,\ p+1\medskip \\ 
2\nu +1,\quad 1%
\end{array}%
\right) ,  \notag
\end{eqnarray}%
\begin{eqnarray}
&&4\mu \left( 2a\beta \right) ^{p}\ \frac{\Gamma \left( 2\nu +1\right) }{%
\Gamma \left( 2\nu +p+1\right) }\ C_{p}  \label{TC} \\
&&\qquad =a\left( \mu +a\kappa \right) ~_{3}F_{2}\left( 
\begin{array}{c}
1-n,\ -p,\ p+1\medskip \\ 
2\nu +1,\quad 1%
\end{array}%
\right)  \notag \\
&&\qquad \quad -a\left( \mu -a\kappa \right) ~_{3}F_{2}\left( 
\begin{array}{c}
-n,\ -p,\ p+1\medskip \\ 
2\nu +1,\quad 1%
\end{array}%
\right) .  \notag
\end{eqnarray}%
The averages of $r^{p}$ for the relativistic hydrogen atom were evaluated in
the late 1930s by Davis \cite{Davis} as a sum of certain three $_{3}F_{2}$
functions. But it has been realized only recently that these series are, in
fact, linearly dependent and related to the Chebyshev polynomials of a
discrete variable \cite{Suslov}. Our formulas here present the final result
(we use the standard definition of the generalized hypergeometric series
throughout the paper \cite{Ba}, \cite{Erd}).\medskip

Analogs of the Nikiforov and Uvarov form of the relativistic radial
functions \cite{Ni:Uv}, \cite{Sus:Trey} are given by%
\begin{eqnarray}
&&4\left( p+1\right) \varepsilon \mu \nu \left( 2a\beta \right) ^{p}\ A_{p}
\label{NUA} \\
&&\quad =a\left( \varepsilon \kappa +\nu \right) \left( 2\left( \varepsilon
\kappa -\nu \right) +p+1\right)  \notag \\
&&\qquad \times \frac{\Gamma \left( 2\nu +p+3\right) }{\Gamma \left( 2\nu
+2\right) }~_{3}F_{2}\left( 
\begin{array}{c}
1-n,\ p+2,\ -p-1\medskip \\ 
2\nu +2,\quad 1%
\end{array}%
\right)  \notag \\
&&\quad -a\left( \varepsilon \kappa -\nu \right) \left( 2\left( \varepsilon
\kappa +\nu \right) +p+1\right)  \notag \\
&&\qquad \times \frac{\Gamma \left( 2\nu +p+1\right) }{\Gamma \left( 2\nu
\right) }~_{3}F_{2}\left( 
\begin{array}{c}
-n,\ p+2,\ -p-1\medskip \\ 
2\nu ,\quad 1%
\end{array}%
\right) ,  \notag
\end{eqnarray}%
\begin{eqnarray}
&&4\mu \nu \left( 2a\beta \right) ^{p}\ B_{p}  \label{NUB} \\
&&\quad =a\left( \varepsilon \kappa +\nu \right) \frac{\Gamma \left( 2\nu
+p+3\right) }{\Gamma \left( 2\nu +2\right) }~_{3}F_{2}\left( 
\begin{array}{c}
1-n,\ p+2,\ -p-1\medskip \\ 
2\nu +2,\quad 1%
\end{array}%
\right)  \notag \\
&&\quad -a\left( \varepsilon \kappa -\nu \right) \frac{\Gamma \left( 2\nu
+p+1\right) }{\Gamma \left( 2\nu \right) }~_{3}F_{2}\left( 
\begin{array}{c}
-n,\ p+2,\ -p-1\medskip \\ 
2\nu ,\quad 1%
\end{array}%
\right) ,  \notag
\end{eqnarray}%
\begin{eqnarray}
&&8\left( p+1\right) \varepsilon \mu ^{2}\nu \left( 2a\beta \right) ^{p}\
C_{p}  \label{NUC} \\
&&\quad =a\left( \varepsilon \kappa +\nu \right) \left( 2\kappa \left(
\varepsilon \kappa -\nu \right) +\left( p+1\right) \left( \kappa
-\varepsilon \nu \right) \right)  \notag \\
&&\qquad \times \frac{\Gamma \left( 2\nu +p+3\right) }{\Gamma \left( 2\nu
+2\right) }~_{3}F_{2}\left( 
\begin{array}{c}
1-n,\ p+2,\ -p-1\medskip \\ 
2\nu +2,\quad 1%
\end{array}%
\right)  \notag \\
&&\quad -a\left( \varepsilon \kappa -\nu \right) \left( 2\kappa \left(
\varepsilon \kappa +\nu \right) +\left( p+1\right) \left( \kappa
+\varepsilon \nu \right) \right)  \notag \\
&&\qquad \times \frac{\Gamma \left( 2\nu +p+1\right) }{\Gamma \left( 2\nu
\right) }~_{3}F_{2}\left( 
\begin{array}{c}
-n,\ p+2,\ -p-1\medskip \\ 
2\nu ,\quad 1%
\end{array}%
\right) .  \notag
\end{eqnarray}%
These representations simplify Eqs. (3.7)--(3.9) of \cite{Suslov} with the
help of the linear relation (\ref{indint1}) (the calculation details are
left to the reader).\medskip\ 

It is important noting in this paper, that formulas (\ref{TA})--(\ref{TC})
and (\ref{NUA})--(\ref{NUC}) provide explicit examples (of inverses) of the
linear transformations (\ref{Trans}) that reduce the original three-term
recurrence relations (\ref{3termA})--(\ref{3termB}) to the difference
equations of the corresponding dual Hahn polynomials in a complete analogy
with the case of the relativistic radial wave functions (see, for example, 
\cite{Ni:Uv} and \cite{Sus:Trey}). One may choose any two of three linearly
dependent integrals $A_{p},$ $B_{p},$ and $C_{p}$ and take the corresponding
renormalized dual Hahn polynomials as $X_{p}$ and $Y_{p}.$\medskip 

For example, by choosing $A_{p}$ and $B_{p}$ as the independent integrals
and introducing 
\begin{eqnarray}
X_{p} &=&{}_{3}F_{2}\left( 
\begin{array}{c}
1-n,\ -p,\ p+1\medskip  \\ 
2\nu +1,\quad 1%
\end{array}%
\right) ,  \label{XY} \\
Y_{p} &=&{}_{3}F_{2}\left( 
\begin{array}{c}
-n,\ \medskip -p,\ p+1 \\ 
2\nu +1,\quad 1%
\end{array}%
\right) ,  \notag
\end{eqnarray}%
from (\ref{TA})--(\ref{TB}) we arrive at the following transformation matrix%
\begin{eqnarray}
T_{p} &=&\frac{\left( 2a\beta \right) ^{p}}{2a^{2}}\frac{\Gamma \left( 2\nu
+1\right) }{\Gamma \left( 2\nu +p+1\right) }  \label{T} \\
&&\times \left( 
\begin{array}{cc}
\dfrac{a\left( 2\kappa +\varepsilon \left( p+1\right) \right) +2\mu }{\mu
+a\kappa } & -\dfrac{a\left( 2\varepsilon \kappa +p+1\right) +2\varepsilon
\mu }{\mu +a\kappa }\medskip  \\ 
-\dfrac{a\left( 2\kappa +\varepsilon \left( p+1\right) \right) -2\mu }{\mu
-a\kappa } & \dfrac{a\left( 2\varepsilon \kappa +p+1\right) -2\varepsilon
\mu }{\mu -a\kappa }%
\end{array}%
\right)   \notag
\end{eqnarray}%
with%
\begin{equation}
\det T_{p}=\left( \left( 2a\beta \right) ^{p}\frac{\Gamma \left( 2\nu
+1\right) }{\Gamma \left( 2\nu +p+1\right) }\right) ^{2}\frac{\mu \left(
p+1\right) }{a\left( \mu ^{2}-a^{2}\kappa ^{2}\right) }.  \label{detT}
\end{equation}%
Then%
\begin{eqnarray}
&&\widetilde{S}_{p}=T_{p}S_{p}T_{p-1}^{-1}=\left( a^{2}p\left( 2\nu
+p\right) \right) ^{-1}  \label{NewSEx} \\
&&\times \left( 
\begin{array}{cc}
-a^{2}p^{2}+2a\varepsilon \mu p-2\left( \mu ^{2}-a^{2}\kappa ^{2}\right)  & 
2\left( \mu ^{2}-a^{2}\kappa ^{2}\right) \medskip  \\ 
-2\left( \mu ^{2}-a^{2}\kappa ^{2}\right)  & a^{2}p^{2}+2a\varepsilon \mu
p+2\left( \mu ^{2}-a^{2}\kappa ^{2}\right) 
\end{array}%
\right)   \notag
\end{eqnarray}%
with the help of the matrix identity (\ref{matrix}) and%
\begin{equation}
\widetilde{\Delta }_{p}=\det \widetilde{S}_{p}=\frac{2\nu -p}{2\nu +p}.
\label{detNewSEx}
\end{equation}%
The new system (\ref{MatSolX}) takes much simplier form%
\begin{eqnarray}
X_{p} &=&-\frac{a^{2}p^{2}-2a\varepsilon \mu p+2\left( \mu ^{2}-a^{2}\kappa
^{2}\right) }{a^{2}p\left( 2\nu +p\right) }\ X_{p-1}  \label{NewX} \\
&&+\frac{2\left( \mu ^{2}-a^{2}\kappa ^{2}\right) }{a^{2}p\left( 2\nu
+p\right) }\ Y_{p-1}  \notag
\end{eqnarray}%
and%
\begin{eqnarray}
Y_{p} &=&-\frac{2\left( \mu ^{2}-a^{2}\kappa ^{2}\right) }{a^{2}p\left( 2\nu
+p\right) }\ X_{p-1}  \label{NewY} \\
&&+\frac{a^{2}p^{2}+2a\varepsilon \mu p+2\left( \mu ^{2}-a^{2}\kappa
^{2}\right) }{a^{2}p\left( 2\nu +p\right) }\ Y_{p-1}  \notag
\end{eqnarray}%
with the initial data%
\begin{eqnarray}
&&\left( 
\begin{array}{c}
X_{0}\medskip  \\ 
Y_{0}%
\end{array}%
\right) =T_{0}\ \left( 
\begin{array}{c}
A_{0}\medskip  \\ 
B_{0}%
\end{array}%
\right)   \label{NewIni} \\
&&\ =\frac{1}{2a^{2}}\left( 
\begin{array}{cc}
\dfrac{a\left( 2\kappa +\varepsilon \right) +2\mu }{\mu +a\kappa } & -\dfrac{%
a\left( 2\varepsilon \kappa +1\right) +2\varepsilon \mu }{\mu +a\kappa }%
\medskip  \\ 
-\dfrac{a\left( 2\kappa +\varepsilon \right) -2\mu }{\mu -a\kappa } & \dfrac{%
a\left( 2\varepsilon \kappa +1\right) -2\varepsilon \mu }{\mu -a\kappa }%
\end{array}%
\right) \ \left( 
\begin{array}{c}
1\medskip  \\ 
\varepsilon 
\end{array}%
\right) =\left( 
\begin{array}{c}
1\medskip  \\ 
1%
\end{array}%
\right) \bigskip .  \notag
\end{eqnarray}%
After this transformation, the three-term recurrence relations (\ref{3termX}%
)--(\ref{3termY}) coincide with the difference equations for the
corresponding special dual Hahn polynomials (\ref{rr05}) (one should use the
spectral identity $\varepsilon \mu =a\left( \nu +n\right) ,$ further
computational details are left to the reader). Our consideration shows how
the relativistic Coulomb expectation values $A_{p}$ and $B_{p}$ can be
independently obtained in their closed forms (\ref{TA})--(\ref{TB}), when
solving the original system (\ref{rra})--(\ref{rrb}) by the methods of the
theory of difference equations developed in the previous section and without
explicit evaluation of the integrals. A striking similarity with the
structure of the radial wave functions provides a guidance in this approach.
A similar analysis of the case (\ref{NUA})--(\ref{NUB}) is left to the
reader.\medskip\ 

On the second hand, our equations (\ref{TA})--(\ref{TC}) and (\ref{NUA})--(%
\ref{NUC}) imply the following linear relations:%
\begin{eqnarray}
&&{}_{3}F_{2}\left( 
\begin{array}{c}
1-n,\ -p,\ p+1\medskip \\ 
2\nu +1,\quad 1%
\end{array}%
\right)  \label{L1} \\
&&\quad =\frac{\left( 2\nu +n\right) \left( 2\nu +p+1\right) \left( 2\nu
+p+2\right) \left( 2n+p+1\right) }{4\nu \left( 2\nu +1\right) \left( \nu
+n\right) \left( p+1\right) }  \notag \\
&&\qquad \qquad \times {}_{3}F_{2}\left( 
\begin{array}{c}
1-n,\ p+2,\ -p-1\medskip \\ 
2\nu +2,\quad 1%
\end{array}%
\right)  \notag \\
&&\qquad -\frac{n\left( 4\nu +2n+p+1\right) }{2\left( \nu +n\right) \left(
p+1\right) }~_{3}F_{2}\left( 
\begin{array}{c}
-n,\ p+2,\ -p-1\medskip \\ 
2\nu ,\quad 1%
\end{array}%
\right)  \notag
\end{eqnarray}%
and%
\begin{eqnarray}
&&{}_{3}F_{2}\left( 
\begin{array}{c}
-n,\ \medskip -p,\ p+1 \\ 
2\nu +1,\quad 1%
\end{array}%
\right)  \label{L2} \\
&&\quad =\frac{n\left( 4\nu +2n-p-1\right) \left( 2\nu +p+1\right) \left(
2\nu +p+2\right) }{4\nu \left( 2\nu +1\right) \left( \nu +n\right) \left(
p+1\right) }~  \notag \\
&&\qquad \qquad \times {}{}_{3}F_{2}\left( 
\begin{array}{c}
1-n,\ p+2,\ -p-1\medskip \\ 
2\nu +2,\quad 1%
\end{array}%
\right)  \notag \\
&&\qquad -\frac{\left( 2\nu +n\right) \left( 2n-p-1\right) }{2\left( \nu
+n\right) \left( p+1\right) }~_{3}F_{2}\left( 
\begin{array}{c}
-n,\ p+2,\ -p-1\medskip \\ 
2\nu ,\quad 1%
\end{array}%
\right)  \notag
\end{eqnarray}%
between two pairs of the generalized hypergeometric series under
consideration. As required, only one dimensionless parameter is involved in
the transformations. Details of these elementary but rather tedious
calculations are left to the reader.\medskip

In addition, from (3.7) of \cite{Suslov} and (\ref{NUA}) of this paper one
gets%
\begin{eqnarray}
&&\dfrac{p\left( p+1\right) }{2\nu +n}{}\ {}_{3}F_{2}\left( 
\begin{array}{c}
1-n,\ \medskip p+1,\ -p \\ 
2\nu +1,\quad 2%
\end{array}%
\right)  \label{L3} \\
&&\quad =\frac{\left( p-2\nu \right) \left( 2\nu +p+1\right) }{2\left( 2\nu
+1\right) \left( \nu +n\right) }~{}{}_{3}F_{2}\left( 
\begin{array}{c}
1-n,\ p+1,\ -p\medskip \\ 
2\nu +2,\quad 1%
\end{array}%
\right)  \notag \\
&&\qquad +\frac{\nu }{\nu +n}~_{3}F_{2}\left( 
\begin{array}{c}
-n,\ p+1,\ -p\medskip \\ 
2\nu ,\quad 1%
\end{array}%
\right) ,  \notag
\end{eqnarray}%
which complements relation (3.12) of \cite{Suslov}:%
\begin{eqnarray}
&&\dfrac{p\left( p+1\right) }{n+2\nu }\ {}_{3}F_{2}\left( 
\begin{array}{c}
1-n,\ -p,\ p\medskip +1 \\ 
2\nu +1,\quad 2%
\end{array}%
\right)  \notag \\
&&\ =\frac{p\left( p+1\right) }{2\nu +1}~{}_{3}F_{2}\left( 
\begin{array}{c}
1-n,\ 1-p,\ p\medskip +2 \\ 
2\nu +2,\quad 2%
\end{array}%
\right)  \notag \\
&&\ ={}_{3}F_{2}\left( 
\begin{array}{c}
-n,\ -p,\ p+1\medskip \\ 
2\nu +1,\quad 1%
\end{array}%
\right) -{}_{3}F_{2}\left( 
\begin{array}{c}
1-n,\ -p,\ p+1\medskip \\ 
2\nu +1,\quad 1%
\end{array}%
\right)  \label{Chebyshev}
\end{eqnarray}%
reproduced here for completeness. One needs to derive transformations (\ref%
{L1})--(\ref{L3}) directly from the advanced theory of generalized
hypergeometric functions \cite{Ba}, \cite{Erd}.\medskip

It is worth noting, in conclusion, that explicit solutions of the systems of
the first order difference equations with variable coefficients are not
widely available in mathematical literature. This is why, it is important to
investigate in detail a remarkable structure of the expectation values
pointed out in this paper for a classical problem of the relativistic
quantum mechanics, such as spectra of high-$Z$ hydrogenlike ions.\medskip

\noindent \textbf{Acknowledgments.\/} I thank Carlos Castillo-Ch\'{a}vez,
Hal Smith and Vladimir Zakharov for valuable discussions and encouragement.

\appendix

\section{Laguerre and Dual Hahn Polynomials}

The Laguerre polynomials are \cite{Erd}, \cite{Ni:Su:Uv}, \cite{Ni:Uv}%
\begin{equation}
L_{m}^{\alpha }\left( x\right) =\frac{\Gamma \left( \alpha +m+1\right) }{%
m!\;\Gamma \left( \alpha +1\right) }\ _{1}F_{1}\left( 
\begin{array}{c}
-m\medskip \\ 
\alpha +1\medskip%
\end{array}%
;\ x\right) .  \label{Laguerre}
\end{equation}%
The dual Hahn polynomials are given by \cite{Ni:Su:Uv}%
\begin{eqnarray}
w_{m}^{\left( c\right) }\left( s\left( s+1\right) ,a,b\right) &=&\frac{%
\left( 1+a-b\right) _{m}\left( 1+a+c\right) _{m}}{m!}  \label{dhahn} \\
&&\ \times \ _{3}F_{2}\left( 
\begin{array}{c}
-m\medskip ,\ a-s,\ a+s+1 \\ 
1+a-b\medskip ,\quad 1+a+c%
\end{array}%
;\ 1\right) .  \notag
\end{eqnarray}%
In (\ref{TA})--(\ref{TC}) and (\ref{NUA})--(\ref{NUC}) of this paper, we are
dealing only with the following special cases: $m=n,n-1$ and $a=b=0,$ $%
c=2\nu ,$ $s=p$ and $a=b=0,$ $c=2\nu \pm 1,$ $s=p+1,$ respectively.\medskip\ 

The difference equation for the dual Hahn polynomials has the form%
\begin{equation}
\sigma \left( s\right) \frac{\Delta }{\nabla x_{1}\left( s\right) }\left( 
\frac{\nabla y\left( s\right) }{\nabla x\left( s\right) }\right) +\tau
\left( s\right) \frac{\Delta y\left( s\right) }{\Delta x\left( s\right) }%
+\lambda _{m}y\left( s\right) =0,  \label{rr01}
\end{equation}%
where $\Delta f\left( s\right) =\nabla f\left( s+1\right) =f\left(
s+1\right) -f\left( s\right) ,$ $x\left( s\right) =s\left( s+1\right) ,$ $%
x_{1}\left( s\right) =x\left( s+1/2\right) ,$ and%
\begin{eqnarray}
&&\sigma \left( s\right) =\left( s-a\right) \left( s+b\right) \left(
s-c\right) ,  \label{rr02} \\
&&\sigma \left( s\right) +\tau \left( s\right) \nabla x_{1}\left( s\right)
=\sigma \left( -s-1\right)  \label{rr03} \\
&&\qquad =\left( a+s+1\right) \left( b-s-1\right) \left( c+s+1\right) , 
\notag \\
&&\lambda _{m}=m.  \label{rr04}
\end{eqnarray}%
It can be rewritten as the three-term recurrence relation%
\begin{eqnarray}
&&\sigma \left( -s-1\right) \nabla x\left( s\right) y\left( s+1\right)
+\sigma \left( s\right) \Delta x\left( s\right) y\left( s-1\right)
\label{rr05} \\
&&\quad +\left( \lambda _{m}\Delta x\left( s\right) \nabla x\left( s\right)
\nabla x_{1}\left( s\right) -\sigma \left( -s-1\right) \nabla x\left(
s\right) -\sigma \left( s\right) \Delta x\left( s\right) \right) y\left(
s\right) =0.  \notag
\end{eqnarray}%
See \cite{Karlin:McGregor61}, \cite{Ko:Sw} and \cite{Ni:Su:Uv} for more
details on the properties of the dual Hahn polynomials.\medskip

\section{Matrix Identity}

The required matrix identity%
\begin{eqnarray}
&&\left( 
\begin{array}{cc}
\dfrac{a\left( 2\kappa +\varepsilon \left( p+1\right) \right) +2\mu }{\mu
+a\kappa } & -\dfrac{a\left( 2\varepsilon \kappa +p+1\right) +2\varepsilon
\mu }{\mu +a\kappa } \\ 
-\dfrac{a\left( 2\kappa +\varepsilon \left( p+1\right) \right) -2\mu }{\mu
-a\kappa } & \dfrac{a\left( 2\varepsilon \kappa +p+1\right) -2\varepsilon
\mu }{\mu -a\kappa }%
\end{array}%
\right)  \label{matrix} \\
&&\times \left( 
\begin{array}{cc}
-p\left( 4\nu ^{2}\varepsilon +2\kappa \left( p+1\right) +\varepsilon
p\left( 2\kappa \varepsilon +p+1\right) \right) & 4\mu ^{2}\left( p+1\right)
+p\left( 2\kappa \varepsilon +p\right) \left( 2\kappa \varepsilon
+p+1\right) \medskip \\ 
-p\left( 4\nu ^{2}+2\kappa \varepsilon \left( 2p+1\right) +\varepsilon
^{2}p\left( p+1\right) \right) & 4\mu ^{2}\varepsilon \left( p+1\right)
+p\left( 2\kappa \varepsilon +p\right) \left( 2\kappa +\varepsilon \left(
p+1\right) \right)%
\end{array}%
\right)  \notag \\
&&\times \left( 
\begin{array}{cc}
\left( \mu +a\kappa \right) \left( a\left( 2\varepsilon \kappa +p\right)
-2\varepsilon \mu \right) & \left( \mu -a\kappa \right) \left( a\left(
2\varepsilon \kappa +p\right) +2\varepsilon \mu \right) \medskip \\ 
\left( \mu +a\kappa \right) \left( a\left( 2\kappa +\varepsilon p\right)
-2\mu \right) & \left( \mu -a\kappa \right) \left( a\left( 2\kappa
+\varepsilon p\right) +2\mu \right)%
\end{array}%
\right)  \notag \\
&&\ =8a^{2}\mu ^{2}\left( p+1\right) \left( 
\begin{array}{cc}
-a^{2}p^{2}+2a\varepsilon \mu p-2\left( \mu ^{2}-a^{2}\kappa ^{2}\right) & 
2\left( \mu ^{2}-a^{2}\kappa ^{2}\right) \medskip \\ 
-2\left( \mu ^{2}-a^{2}\kappa ^{2}\right) & a^{2}p^{2}+2a\varepsilon \mu
p+2\left( \mu ^{2}-a^{2}\kappa ^{2}\right)%
\end{array}%
\right) ,  \notag
\end{eqnarray}%
provided $a^{2}=1-\varepsilon ^{2}$ and $\mu ^{2}=\kappa ^{2}-\nu ^{2},$ can
be verified with the help of a computer algebra system.

\end{document}